# True Alternating Current Scanning Tunneling Microscope (ACSTM): tunneling on insulators

Marcel J. Rost*

*Huygens-Kamerlingh Onnes Laboratory, Leiden University,*

*Niels Bohrweg 2, 2333 CA Leiden, The Netherlands*

## Abstract

Scanning Tunneling Microscopy (STM) has revolutionized our atomic scale understanding of surfaces and accelerated progress in nanotechnology. This technique, however, is restricted to metal or semiconducting samples, as it requires a tiny current to stabilize the tip-sample distance with atomic scale precision. We developed a new imaging and feedback method that relies on true alternating current (AC) without any direct current (DC) component. This technique does not only enable the imaging on non-conducting surfaces with atomic step resolution, like (thin) glass and oxides, it provides also access to high-frequency electronic signal coming from the sample. We demonstrate that it is possible to measure on 25nm thick silicon oxide with 10 MHz tunneling current.

*Electronic address: rost@physics.leidenuniv.nl

Although the potential of ACSTM and high frequency spectroscopy was recognized as early as 1989 [1–3], it has yet to be fully realized. Despite subsequent progress [4–7], no STM has been reported that operates under true ACSTM conditions without a DC component in the feedback controlling tip–sample distance. Consequently, STM has not been used to image true insulators with atomic (step) resolution. Instead, high frequency or microwave signals are typically superimposed via a bias-T onto a conventional DC feedback loop, enabling access to high frequency responses while maintaining standard control. This approach has enabled impressive time resolutions down to the picosecond range applying high MHz to THz frequencies [8–12], yielding insights into various (quantum) material properties such as ferroelectric domains, noise, Kondo effects, spin polarized currents, vortex dynamics, and spin relaxation [13–19]. Nevertheless, it still relies on (semi)conducting samples enabling tip height control.

Our aim is realizing tunneling under true, high frequency ACSTM conditions thereby ultimately measuring on insulators with atomic (step) resolution. For example, at 1 GHz and a tunneling current of 100 pA, typical for STM, tunnelling would naively involve only *one* single (or a few) electrons oscillating in the AC field, shuttling between tip and sample. This could enable STM measurements on non-conducting materials, since a small number of excess surface electrons is always present due to static charging. The idea is that an AC only STM could capture and exploit this charge for feedback and imaging. This also requires a fully AC based approach mechanism.

Inspired by early work by Weiss [2, 3] and Schmidt [20], who used nonlinear mixing of two high frequency AC signals for feedback, we set out to realize true ACSTM. A key obstacle at high frequencies is the unavoidable stray tip–sample capacitance, which acts as a short and generates a current that overwhelms the tunneling signal. From our prior work approaching an STM blindly using this capacitive current [21], we knew its magnitude and that active suppression or compensation was essential. The problem worsens with increasing frequency and decreasing tip–sample distance, which increases the capacitance. As illustrated in Fig. 1, the intrinsic parallel capacitance $C_{\mathrm{STM}}$, with impedance

$$Z_{C_{\mathrm{STM}}} = 1/(i\omega C_{\mathrm{STM}}), \tag{1}$$

($\omega$ is the angular frequency) produces a current that dominates over the true tunneling current, given typical tunneling resistances of 1 GΩ. This challenge is threefold: the atomic scale gap yields a relatively large capacitance, atomic distance variations cause large impedance changes, and the impedance decreases with frequency, making high-frequency operation increasingly difficult.

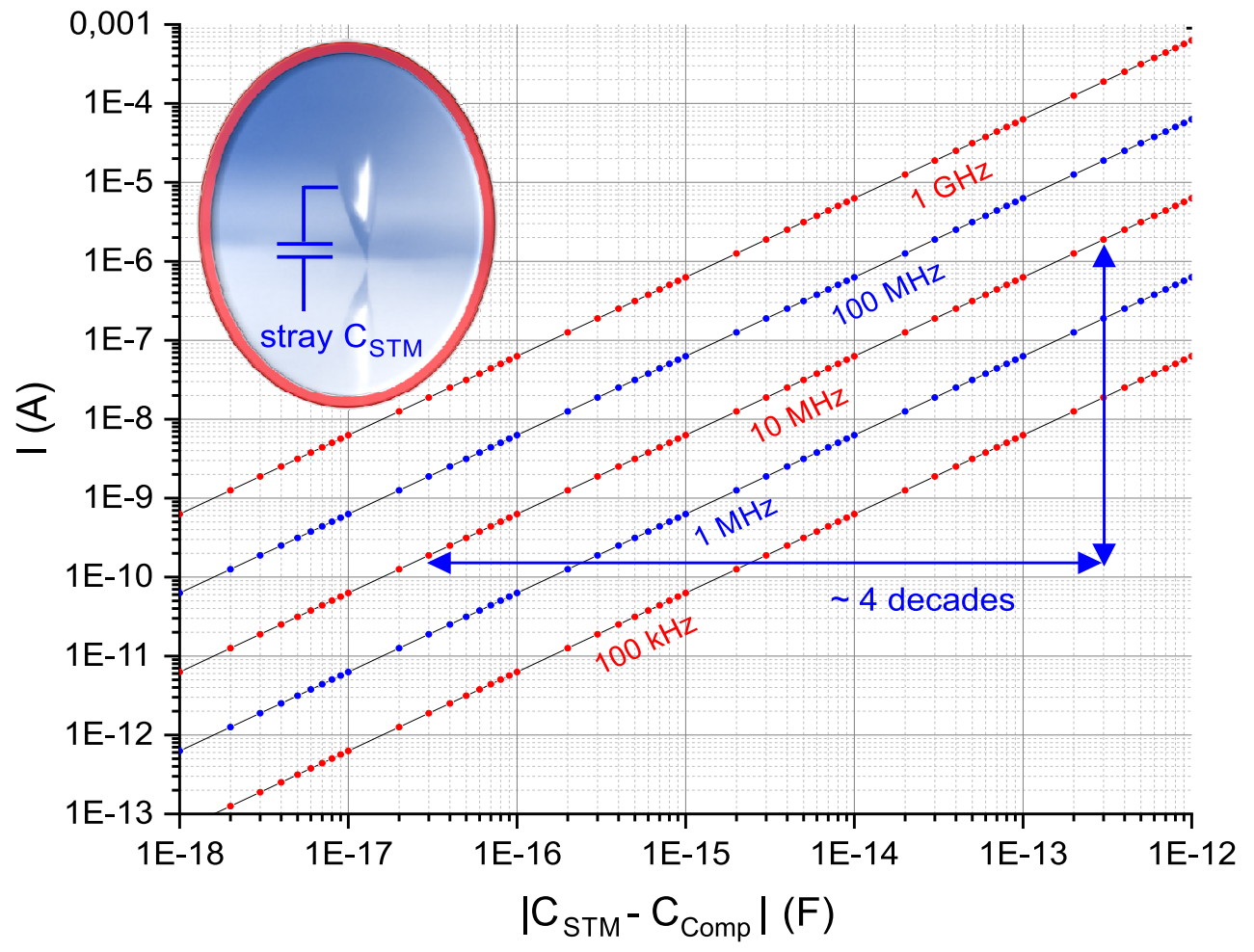

FIG. 1: **Undesired Current** due to the parallel, intrinsic tip-sample capacitance, $C_{\mathrm{STM}}$, at 100 mV excitation voltage as a function of frequency. Without compensation, $C_{\mathrm{Comp}} = 0F$, this current is typically 2 $\mu A$ at 10 MHz [21], as $C_{\mathrm{STM}}$ is in the order of a few 100 $fF$ [21]. It is possible to suppress this current with 4 decades by applying our compensation circuit, thereby enabling the detection of tunneling currents as low as 200 pA.

Various passive and active methods exist to suppress the undesired tip–sample capacitance, including resonant matching circuits, resonant amplifiers, compensating feedback, Wien bridges, phase shifters, (notch) filters, counter current sources, and stub tuners [22]. However, these approaches typically achieve suppressions of only 2.5 orders of magnitude [5, 23]. We therefore introduce a novel scheme (Fig. 2) that also offers a key practical advantage: aside from the STM transfer function, it is broadband, operating from 200 kHz to 1.2 GHz, enabling frequency tuning beyond existing solutions. The circuit functions as a nulling bridge with its signal amplified by a preamplifier, providing high sensitivity for tuning the compensation capacitance. This tuning is performed after the optical coarse approach, before switching to the automatic, fine approach for reaching the tunneling regime. As the tip–sample distance decreases during this process, bridge detuning can occur, leading to false tunneling detection and premature termination of the approach routine. Since no tip-crash occurs, simple retuning suffices. Activating a Phase-Locked-Loop (PLL, Fig. 2) allows continuous tracking of the optimal compensation during approach, eliminating the need for retuning. This scheme is analogous to frequency-modulation based non-contact Atomic Force Microscopy (NC-AFM) [24].

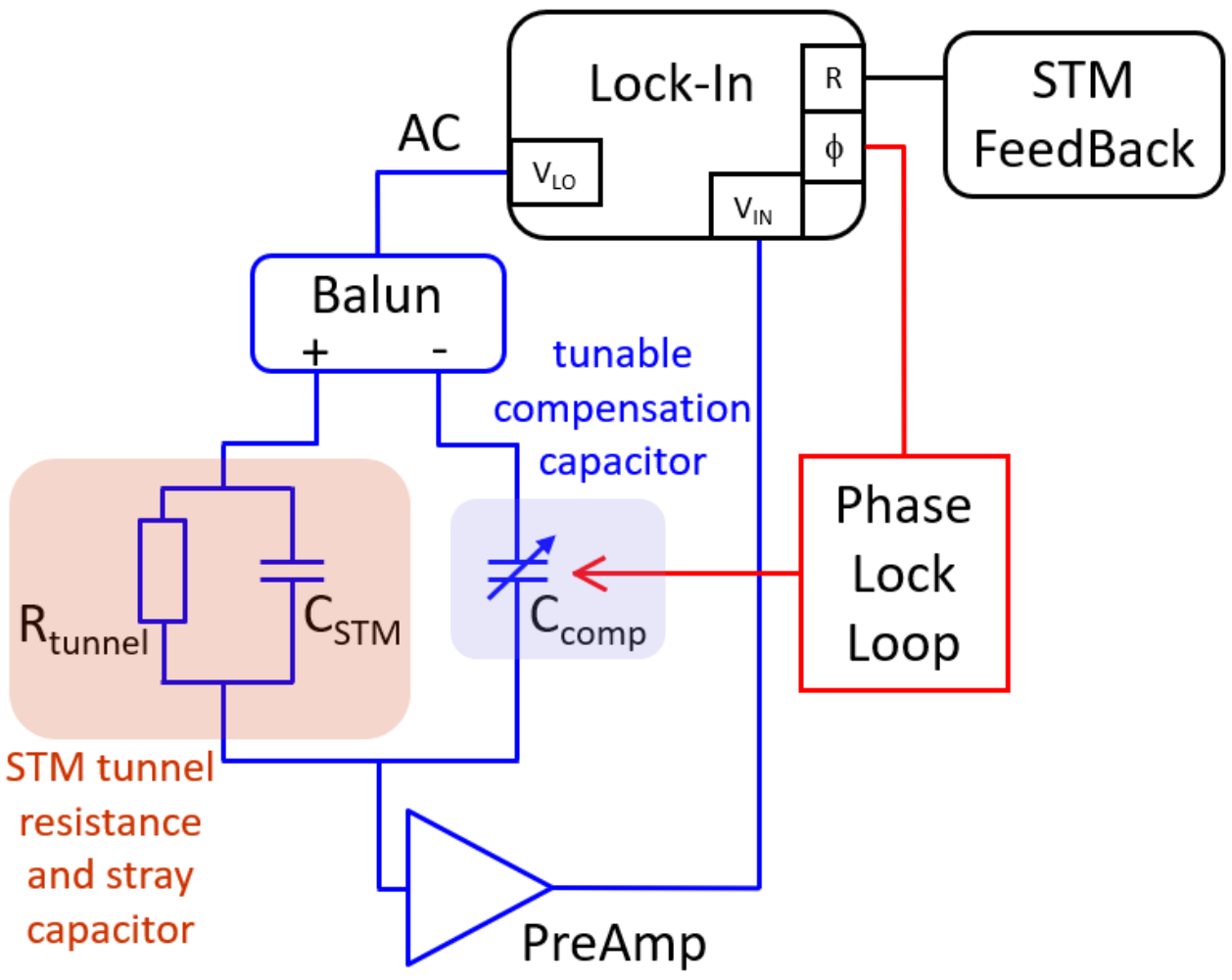

FIG. 2: **Compensation Circuit:** The AC reference output of the LockIn amplifier is converted to a fully identical positive and negative signal via a Balun, from which the one branch feeds the STM, whereas the other one runs via a tunable compensation capacitance. The currents of both branches are converted via a true AC transimpedance preamplifier (capacitive input stage) into a voltage that is demodulated via the LockIn. The amplitude of this signal, thus the rectified tunneling current, is used for the feedback of the STM controller, while the phase can be used to track the optimum compensation via the tunable compensation capacitance $C_{Comp}$ using a Phase-Lock-Loop (PLL).

We validate the technique in three steps: (i) we demonstrate atomic height resolution on a metal surface and directly compare it to a standard STM image taken at the same location; (ii) we confirm that the AC signal is a true tunneling current by showing its exponential distance dependence; and (iii) we demonstrate measurements on 25 nm thick silicon-oxide.

Before presenting Au(111) data, it is important to note that metallic surfaces can in principle also be imaged in AC mode via the capacitance channel, as done in Scanning Capacitance Microscopy (SCM), introduced in 1985 [25] and further refined to attofarad sensitivity [26–29]. However, unlike STM, SCM cannot achieve atomic resolution: its fun-damental height and lateral resolution is limited to around ∼2 nm [30, 31]. This limitation arises from the signal dependence on the tip–sample distance Z: the capacitance typically scales as 1/Z [21], whereas the tunneling current varies exponentially as $\exp(-1.025\sqrt{\phi} \times Z)$, in which Z is measured in Å and $\phi$ is the vacuum work function. Therefore, to demonstrate that our method relies on genuine tunneling rather than capacitive contrast, we must show not only imaging capability but true atomic (step) resolution.

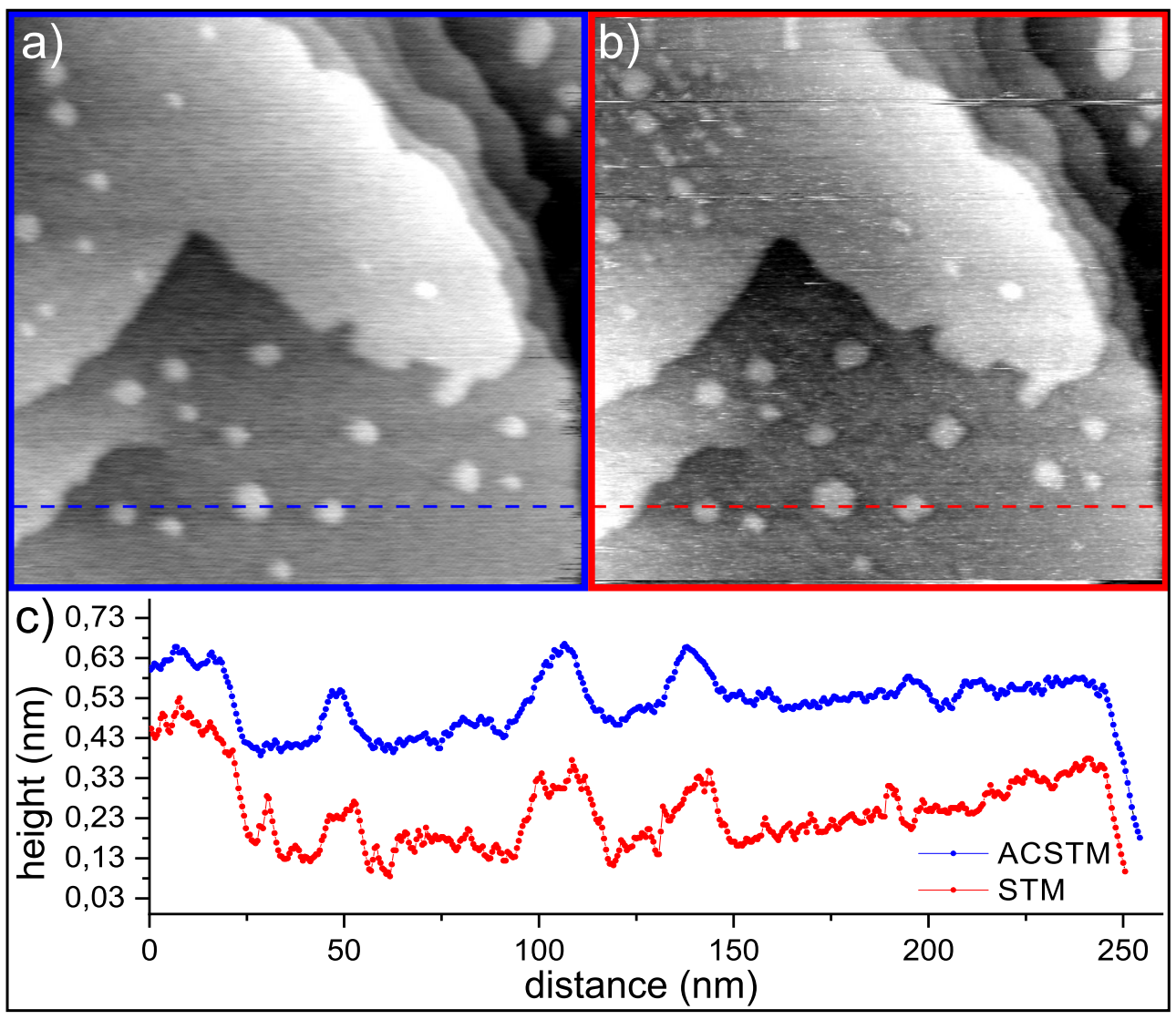

FIG. 3: **Atomic Height Resolution on Gold:** Direct comparison on the very same place of the sample between ACSTM (blue) and standard STM (red). Individual adatom islands as well as mono-atomic height steps are resolved. This ACSTM resolution exceeds the theoretical limit of a Scanning Capacitance Microscope (SCM) with a factor of at least 12 indicating that a different contrast mechanism, like tunneling, is responsible. ACSTM: 10 MHz, $U_{sample}$ = 100 mV, $I_t \simeq 6$ nA, frame time = 58s; STM: $U_{sample}$ = -700 mV, $I_t \simeq 70$ pA; frame time = 170s. The height lines are offset in Z to enable a better comparison.

We assess the resolution on 200 nm thick, highly textured Au(111) films on mica that had been stored for about one year. Such samples are known to host sulfur related adsorbates that pull individual Au atoms from terraces and form adatom–sulfur islands. In conventional STM, the strong tip interaction can activate diffusion and Ostwald ripening of these weakly bound complexes. Figure 3 directly compares ACSTM and conventional STM images taken at the same location. Aside from ripening effects, both images show identical atomic features: individual adatom islands and atomic steps. This confirms true AC tunneling, as the achieved resolution exceeds the physical limit of SCM by roughly an order of magnitude. Any residual DC contribution can be excluded, since our preamplifier has a purely capacitive AC input stage with a bandwidth of 10 kHz to 1.2 GHz (see also Supplementary). Notably, the ACSTM image exhibits lower noise and higher stability, although prior conventional STM imaging, possible tip changes, surface cleaning, and different acquisition times may also contribute.

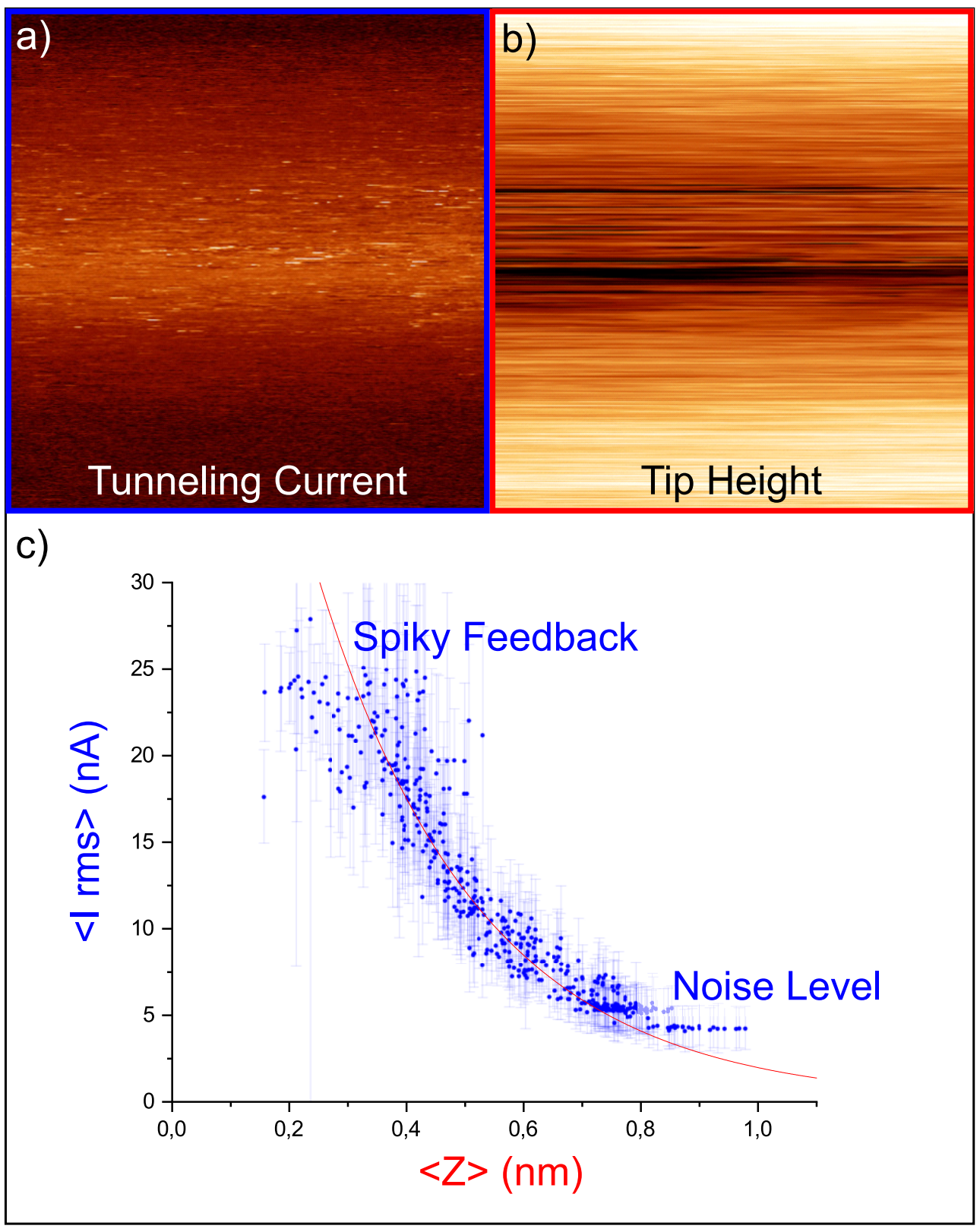

FIG. 4: **Exponential Tunneling Current:** Standing on the surface and not scanning, we recorded with fully active feedback (a) the tunneling current and (b) the simultaneously measured height, while continuously changing the tunneling current setpoint. Averaging these values per line and plotting against each other delivers the graph in (c), in which the error bars are the standard deviation of the tunneling current. The exponential fit (red) delivers a work function of $\phi \approx 0.14eV$.

The atomic step resolution, the real (in-phase) nature of the 10 MHz AC current, with a phase distinct from purely capacitive signals [60], and the absence of contact with the surface (no preamplifier saturation) all indicate a genuine tunneling current. To further confirm this, we measured its dependence on the tip–sample distance Z. Because room temperature, ambient operation is less stable than low-temperature vacuum STM, we obtained the I–Z characteristic in active height feedback by varying the tunneling setpoint while remaining stationary (not scanning) on the surface.

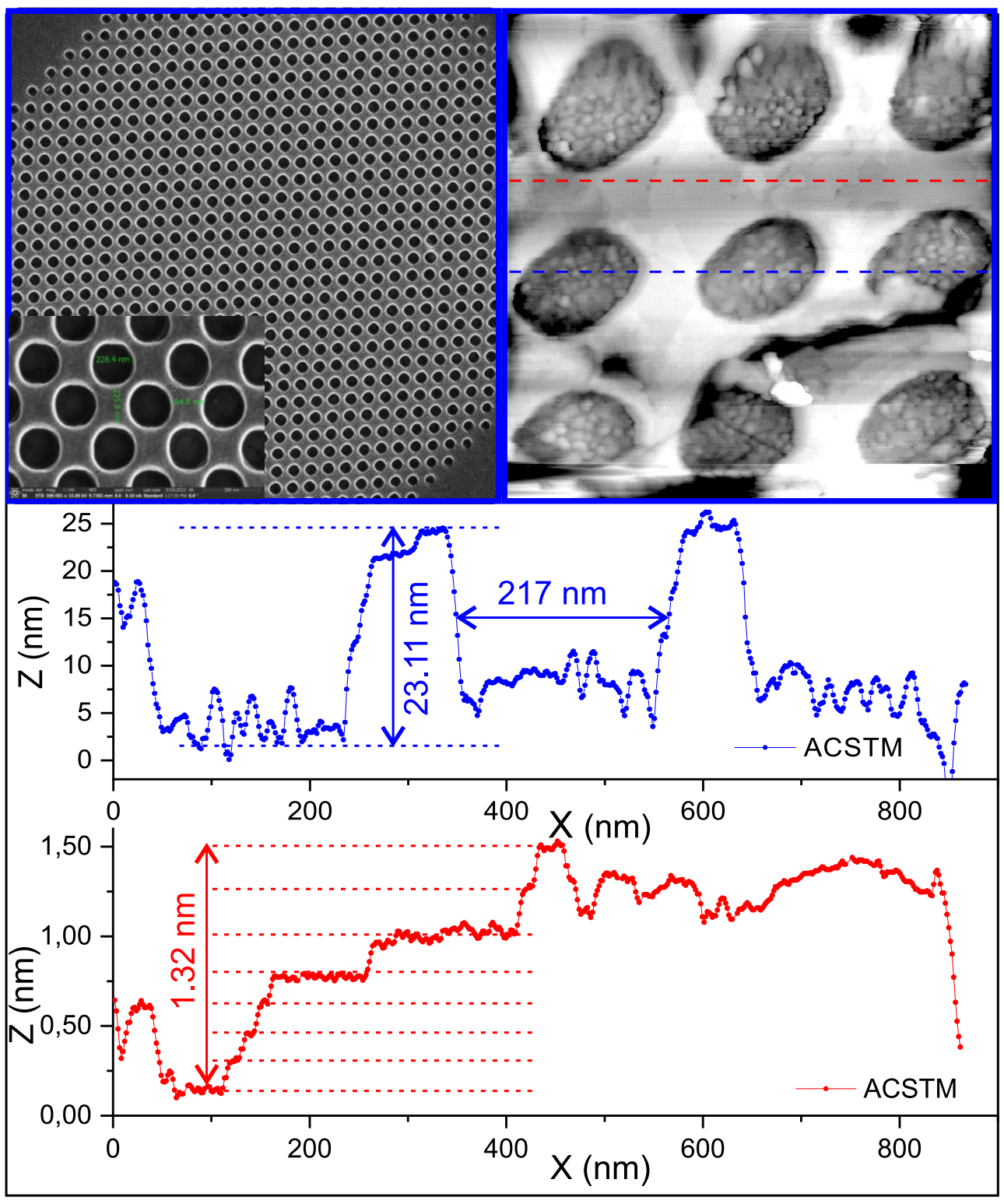

FIG. 5: **ACSTM on $SiO_2$:** we deposited 25nm thick $SiO_2$ on our gold sample and used nano-lithography and reactive ion etching to create a *square* like pattern with 228 nm large holes, in which the bare gold surface is exposed. Stable ACSTM imaging demonstrates that it is possible to measure on 25 nm thick insulators using 10 MHz, $U_{sample}$ = 100 mV, $I_t \simeq 12$ nA, and a frame time = 3772 s. The red height-line is measured on the $SiO_2$ film resolving individual steps with a height of $1.9 \pm 0.1$ Å, which coincides with the theoretical value for $\alpha$−$SiO_2$(0001) [39]. We verified the existence of single crystalline quartz with a $\Theta - 2\Theta$ XRay measurement [40].

Figure4a shows the measured tunneling current with the corresponding Z response in Fig. 4b. Line averaging yields the $< I >$ versus $< Z >$ relation shown in Fig.4c. At low currents we reach the noise floor of the tuned circuit, while at high currents the tip is very close to the surface, naturally leading to instabilities, increased noise, and spikes, visible also in Fig.4a. Fitting the data with an exponential decay (zero offset) yields an effective work function of ∼0.14 eV, about 38x smaller than typical values for clean gold (∼ 5.4 eV). Several factors likely contribute to this reduction. Under ambient conditions, measured work functions are substantially lower (∼1 eV for fresh gold) and can decrease further to several tenths of eV with the duration of air exposure, surface adsorbates, and water condensation at hydrophilic metal surfaces [32, 33]. In addition, our aged sample contains significant sulfur/thiol contamination, evidenced by the adatom islands, which is known to reduce the work function in vacuum by up to ∼1.4 eV [34]. The resulting water meniscus can form an electrochemical environment with double layers and adsorbates, lowering the barrier further and potentially modifying the tunneling behavior beyond a simple exponential [35, 36]. Finally, the modulated barrier inherent to high frequency AC tunneling may also affect the effective work function, an effect beyond the scope of the present work.

After demonstrating atomic (height) resolution, we extended the technique to thicker oxide films. Schneider showed in 2002 that STM can achieve atomic resolution on MgO layers grown on Ag, and could image even the buried Ag atoms through the oxide, but only up to ∼3 MgO layers due to the significantly reduced Local Density of States (LDOS) [37]. Truly thick, insulating films therefore remain inaccessible to conventional STM. Using our NanoLab facilities, we sputter deposited ∼25 nm of $SiO_2$ onto Au films, and then used nanolithography and reactive ion etching to create a distorted square array of circular holes (∼228 nm diameter; Fig.5). During growth, a rough, thin Au–silicide interfacial layer forms before pure $SiO_2$ grows on top. Etching removes both Si and O, exposing ∼6 nm high Au nano-islands within the holes that are remainders of the nanoscale Au–silicide interface [38]. Measured from the top of the nanoislands, this results in a ∼20 nm thick $SiO_2$ "sieve" firmly attached to gold. The ACSTM image in Fig.5 clearly resolves this structure and demonstrates tunneling not only on the exposed Au inside the holes, but also on top of and across the 25nm thick $SiO_2$. Line profiles confirm a hole diameters of 217 nm and an $SiO_2$ thickness of 22.5 nm. Applying 100 mV excitation at 10 MHz, we received a tunneling current of ∼12 nA, corresponding to ∼7500 electrons oscillating simultaneously. A typical Fourier spectrum and corresponding time traces of the preamplified signal are shown in the Supplementary. Remarkably, individual atomic steps and layers are visible even on the $SiO_2$, which cannot be achieved with SCM, again indicating tunneling as the contrast mechanism.

The red height line in Fig. 5, measured on the oxide, shows atomic steps with a height of (1.9 ± 0.1) Å, which coincides with the theoretical value for $\alpha$ -$SiO_2$(0001). We have further evidence that the deposited $SiO_2$ forms large patches of single crystalline quartz, $SiO_2$(0001) [40]. Additional ACSTM images on the oxide are presented in the Supplementary.

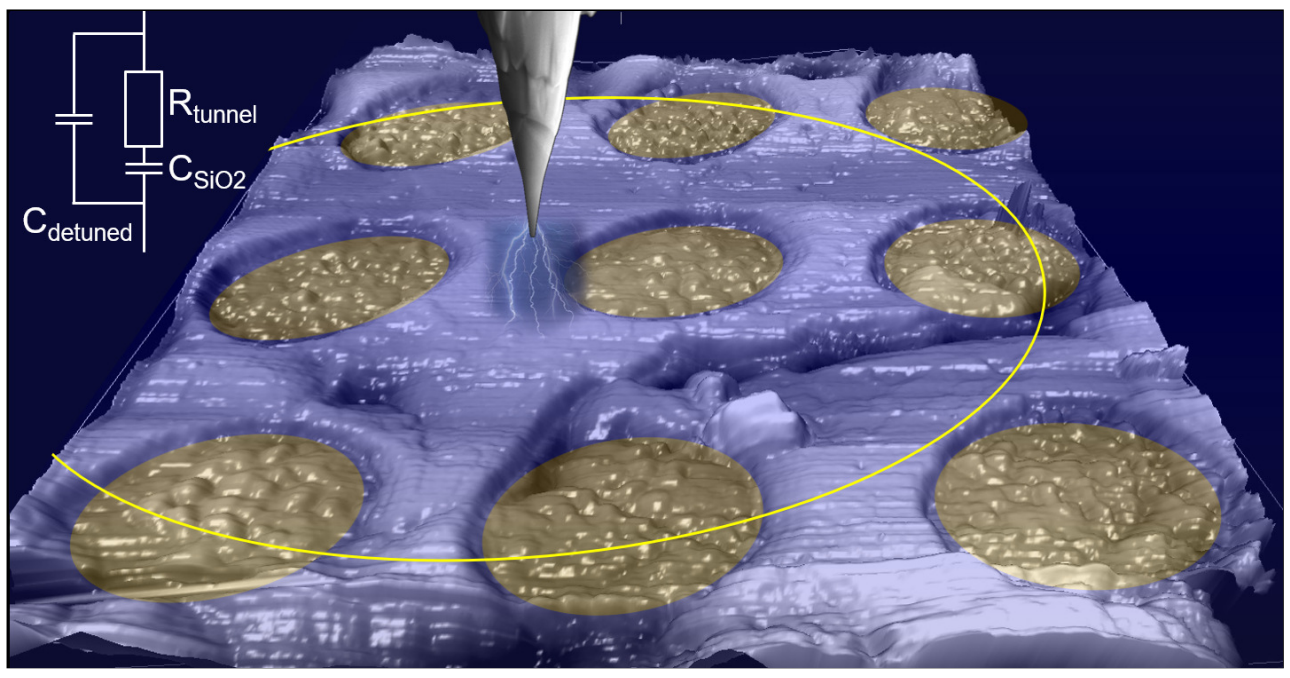

FIG. 6: **Footprint of Charge Spreading:** Assuming that the $SiO_2$ layer works as a capacitance in series with the tip, like drawn in the upper left scheme, one can calculate the required charge spreading on the $SiO_2$ surface that eventually would allow tunneling to occur (see text for more details). The corresponding value of r = 530 nm is indicated to scale by the yellow circle. Note that there will be always eight conducting gold holes within this circle, which makes the required charge spreading significantly less stringent.

An open and intriguing question is the mechanism enabling tunneling through the 25 nm thick $SiO_2$ layer. Quantitative estimates make this result even more surprising. If one models the $SiO_2$ as a simple parallel plate capacitor, with the Au substrate as one electrode and the insulating surface as the other, the required surface charging can be estimated. Using the dielectric constant of quartz ($\epsilon$=4.6) and assuming a generously large electron footprint under the tip (r=10 nm), the potential needed to sustain tunneling would be ~2070 V; clearly unphysical and incompatible with tunneling. This is further supported by impedance considerations: the $SiO_2$ layer impedance ($Z_{C_{SiO_2}} = 2.7 \times 10^{10}\Omega$) vastly exceeds that of the tunneling junction ($Z_{tunnel} = 8.3 \times 10^{6}\Omega$; Fig.6), which would imply that only two of all the measured electrons would flow over the tip branch. All other should flow over the detuned capacitance branch [61]. These values show that simple capacitive charging cannot explain our observations, and that one or more additional physical mechanisms must be involved, which we discuss qualitatively below.

Although quartz is generally an excellent insulator ($10^{-17}$ to $10^{-18}$ $S/m$ [42, 43]), its electrical conductivity is strongly anisotropic and can increase dramatically from $5 \times 10^{-15}$ to $2\times10^{-6}$ $S/m$ in $\alpha$-quartz depending on crystal orientation [44]. A first possible mechanism is therefore finite conduction through the quartz layer. Similar behavior has been reported for STM tunneling through CaO(001) films up to 50 ML (12.3 nm) thick [45]. There, poor screening of the tip induced electric field leads to strong band bending, and although the films are too thick for direct tunneling, electrons are injected into the oxide conduction band and traverse the film incoherently, likely via phonon assisted hopping rather than ballistic transport [45–47]. Comparable mechanisms may contribute in our case.

Additional effects likely arise from the quartz surface and the water condensate interface. Injected surface electrons can be efficiently screened by the polaronic response of $SiO_2$, surface adsorbates, and the electric double layer formed by water dipoles and ions, substantially reducing the effective energy cost. While intrinsic surface conductance of quartz is very low ($\sim 10^{-9}$ S[48, 49]), it is not strictly insulating: humidity, adsorbates, defects, impurities, and surface treatments can raise it to the low $10^{-6}$ S range[49], and ion containing water films can reach $\sim 10^{-4}$ S[50]. Under such conditions, lateral spreading of surface charge, driven by strong Coulomb repulsion, must also be taken into account.

There is precedent for tunneling on nominally insulating surfaces mediated by thin water layers. Using ultralow DC tunneling currents (<1 pA), conventional STM imaging of DNA on mica, both strongly hydrophilic and fully insulating, has been demonstrated [51]. This was attributed to the unexpectedly high lateral conductivity of adsorbed water films, up to five orders of magnitude greater than bulk water, driven by proton transport via the Grotthuss mechanism (hydrogen bond–mediated proton hopping) [52, 53]. Given the presence of a water condensate in our system, a similar contribution is likely.

Further insight comes from AFM studies of charge spreading on insulating surfaces in vacuum [54]. There, $\sim$300 electrons injected without contact onto iron oxide at 4.7 K formed polarons and, upon annealing to 24 K, spread isotropically over >2 $\mu$m. This demonstrates significant surface conductivity on insulators, driven by Coulomb repulsion and variations in the tip-induced electric field; mechanisms that may also be relevant here.

We can contextualize our results by estimating the surface charge spreading radius required to match the experimental conditions. Tunneling becomes feasible for a footprint radius of roughly r $\sim$ 530 nm, at which the quartz impedance drops to $Z_{C_{SiO_2}} = 9.8 \times 10^6\Omega$, below the tunneling impedance $Z_{tunnel} = 5.7 \times 10^7\Omega$. At this scale, the surface potential would be only $\sim$0.1 V, consistent with our applied AC amplitude. About 6400 electrons would contribute to tunneling, and roughly 1100 (of the 7500 )electrons would flow through the detuned side branch. The required spreading is already smaller than in the referenced precedence and would be sufficient by itself to explain our observations (Fig.6). Moreover, the presence of gold holes within this footprint and a conductive adsorbed water layer would further reduce the necessary spreading radius. While alternative interpretations cannot be excluded, these estimates provide a plausible explanation. Definitive identification of the mechanism will require additional studies, particularly low temperature vacuum STM experiments.

On the technical side, several improvements are already in place. Adding a phase shifter to one branch and an attenuator to the other enables operation at lower currents or higher frequencies; frequencies of 5x to 10x higher should be readily achievable. Charge spreading should remain unproblematic, since the injected charge decreases linearly with increasing frequency at constant tunneling current.

The technique provides direct access to the high-frequency tunneling current, opening applications in electronic noise measurements and (low temperature) electronic and materials characterization, such as superconductivity and ferroelectrics. It also enables studies of biological and catalytic systems, including metals on (thin) oxides, with simultaneous topographic and electronic characterization. True ACSTM has the potential to replace conductive AFM–based methods (cAFM), from Scanning Impedance Microscopy (SIM) to Microwave Impedance Microscopy (MIM), by combining feedback and multifrequency operation within the same electronics. Moreover, the feedback scheme can be integrated into high-frequency pump–probe, Electron-Spin-Resonance STMs, microwave, and noise optimized STMs without a bias tee, while extending accessible materials to thicker oxides and nonconductors [55–59].

High-speed STM could be pushed beyond video rate imaging, with additional opportunities inspired by SCM concepts. Finally, demonstrating tunneling through 25 nm thick insulators suggests that even thicker insulators can be accessed at higher frequencies, thereby ultimately entering a new regime, where fundamental questions arise about single-electron tunnelling and charge distribution on insulators.

# I. SUPPLEMENTARY

## A. High Frequency Signal

Figures S1 and S2 show the Fourier Transformation of the high-frequency 10 MHz AC-STM signal measured directly behind the preamplifier (a) as well as six time snap-shots (and their running integrals) of the AC signal after tuning the capacitance compensation. Using an HF2LI Lock-in from Zurich Instruments, the Fourier Transformation has been obtained with a running average of 256 samples. The according scope graphs (b-g) have been obtained within two minutes during this measurement. The Fourier Transformation has been obtained at the time of the first time trace, while all other Fourier Transformations deviate insignificantly. However, the variation in the time traces demonstrate the sensitivity to electrical interference signals, which is the reason for the difficulty of these measurements. This is also reflected in the graphs of the running integrals, which show not only electrical interference but also the drifting of the signal: the limited bandwidth of the complete Z-feedback loop, operating in the 10s of kHz, is unable to adjust the set-point height within $\mu s$ leading to the build up of an offset.

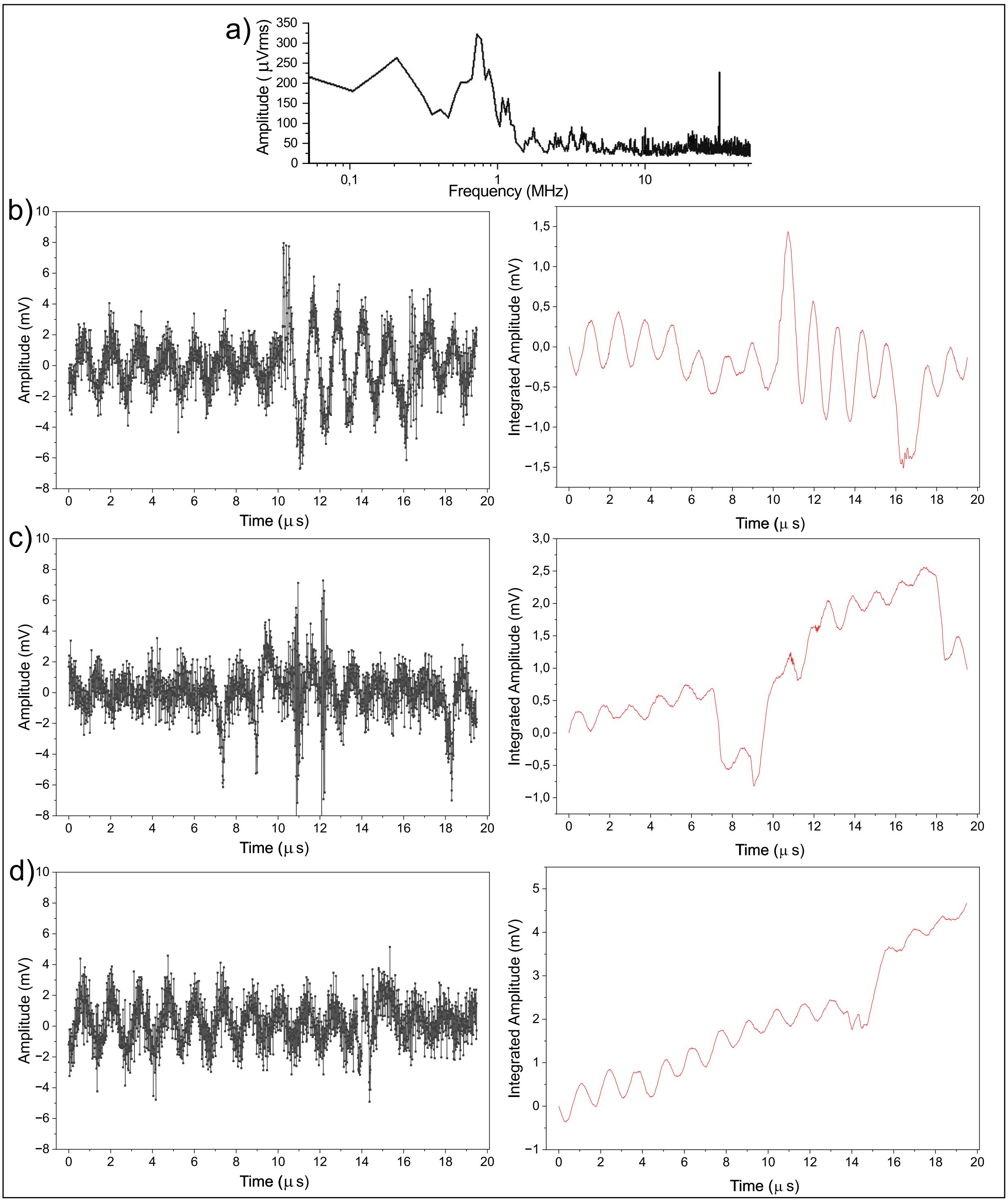

FIG. S1: **The high-frequency 10 MHz ACSTM signal** obtained directly behind the preamplifier: (a) Fourier Transformation measured with a running averge of 256 samples, and (b-g) time trace snap-shots (and their running integrals) of the bare signal allowing direct access to the high-frequency signal coming from the sample. The tiny peak in the Fourier Transform above the noise at 10 MHz corresponds to around 15 nA and demonstrates that we measure on the edge of what is possible.

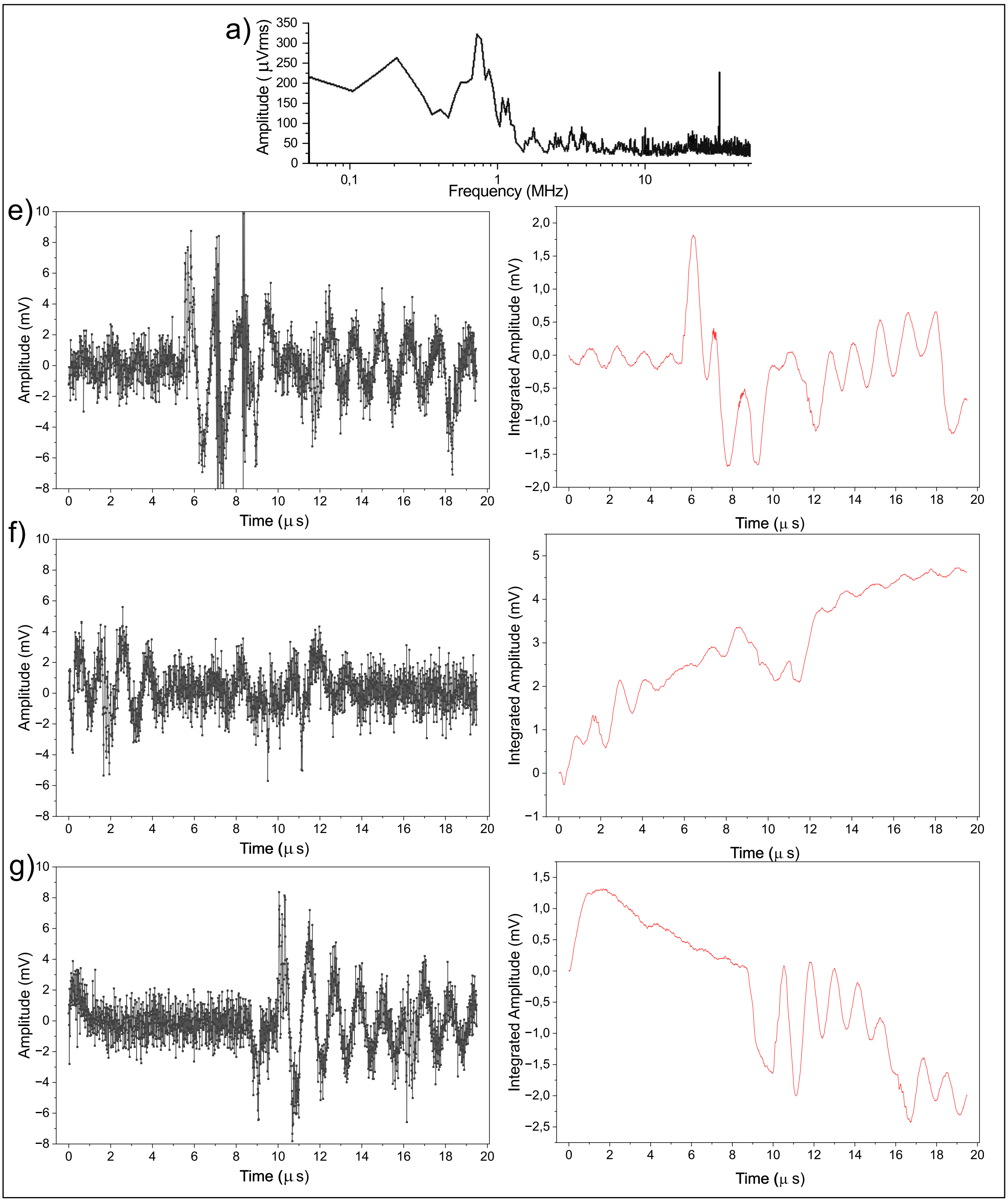

FIG. S2: **The high-frequency 10 MHz ACSTM signal** obtained directly behind the preamplifier: (a) Fourier Transformation measured with a running averge of 256 samples, and (b-g) time trace snap-shots (and their running integrals) of the bare signal allowing direct access to the high-frequency signal coming from the sample. The tiny peak in the Fourier Transform above the noise at 10 MHz corresponds to around 15 nA and demonstrates that we measure on the edge of what is possible.

## B. Preamplifier, Bandwidth, and DC-Block

Standard STM preamplifiers and feedback schemes are designed to detect DC tunnelling currents and typically operate from DC up to several kHz. Thus, when the tip is stationary above the surface, a conventional STM measures a finite DC tunnelling current (neglecting electronic and mechanical noise).

In contrast, our ACSTM electronics is designed to reject the DC component entirely. The preamplifier used in our measurements (FEMTO HSA-X-1-40) has a capacitive input stage with a specified bandwidth of 10 kHz to 1.2 GHz. Since the impedance of a capacitor, $Z = 1/(i\omega C)$, diverges as $\omega \to 0$, the capacitive input stage blocks any DC current. In addition, the bias DC-path at the input stage of the preamplifier acts not only as a short to GND for DC signals but provides also a high-pass response. Consequently, feedback action in our ACSTM case relies on signals above 10KHz without any DC component, while standard STM technology uses exactly the bandwidth from DC up to several kHz.

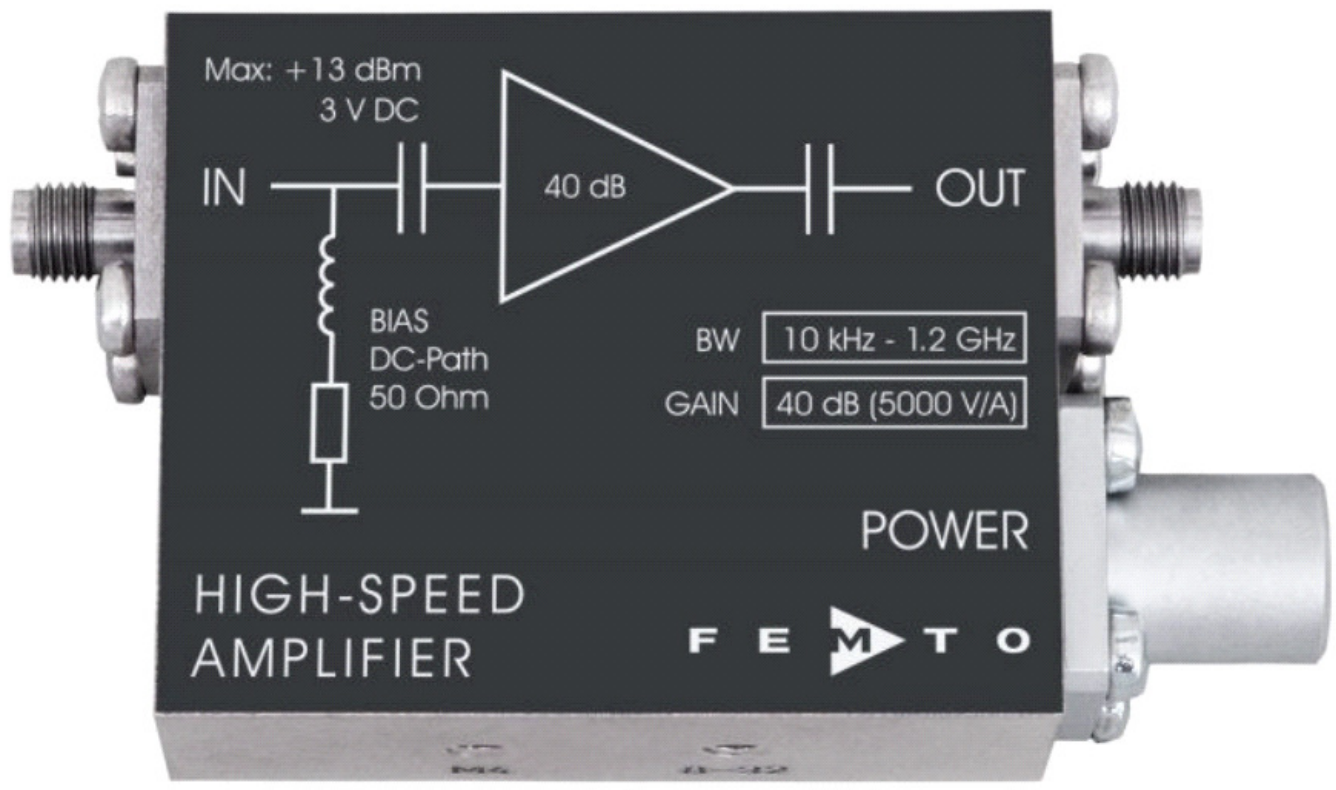

FIG. S3: **Electronic Scheme** of the HSA-X-1-40 from Femto.de.

## C. ACSTM Images

The presented ACSTM images in the main part of the paper are the clearest and most relevant images for presentation demonstrating the achievable image quality and atomic-step resolution. In total more than 65 measurements have been performed, of which 54 measurements in ACSTM mode, and 9 measurements on the specifically nanofabricated SiO2/Au(111) sample. During these measurements, 62 images were obtained in ACSTM mode on the SiO2/Au samples, although the majority do not exhibit sufficient image quality or resolution to be suitable for presentation.

To provide more proof for the achievable resolution on the $SiO_2$ surface, we present here a additional images obtained from our measurements.

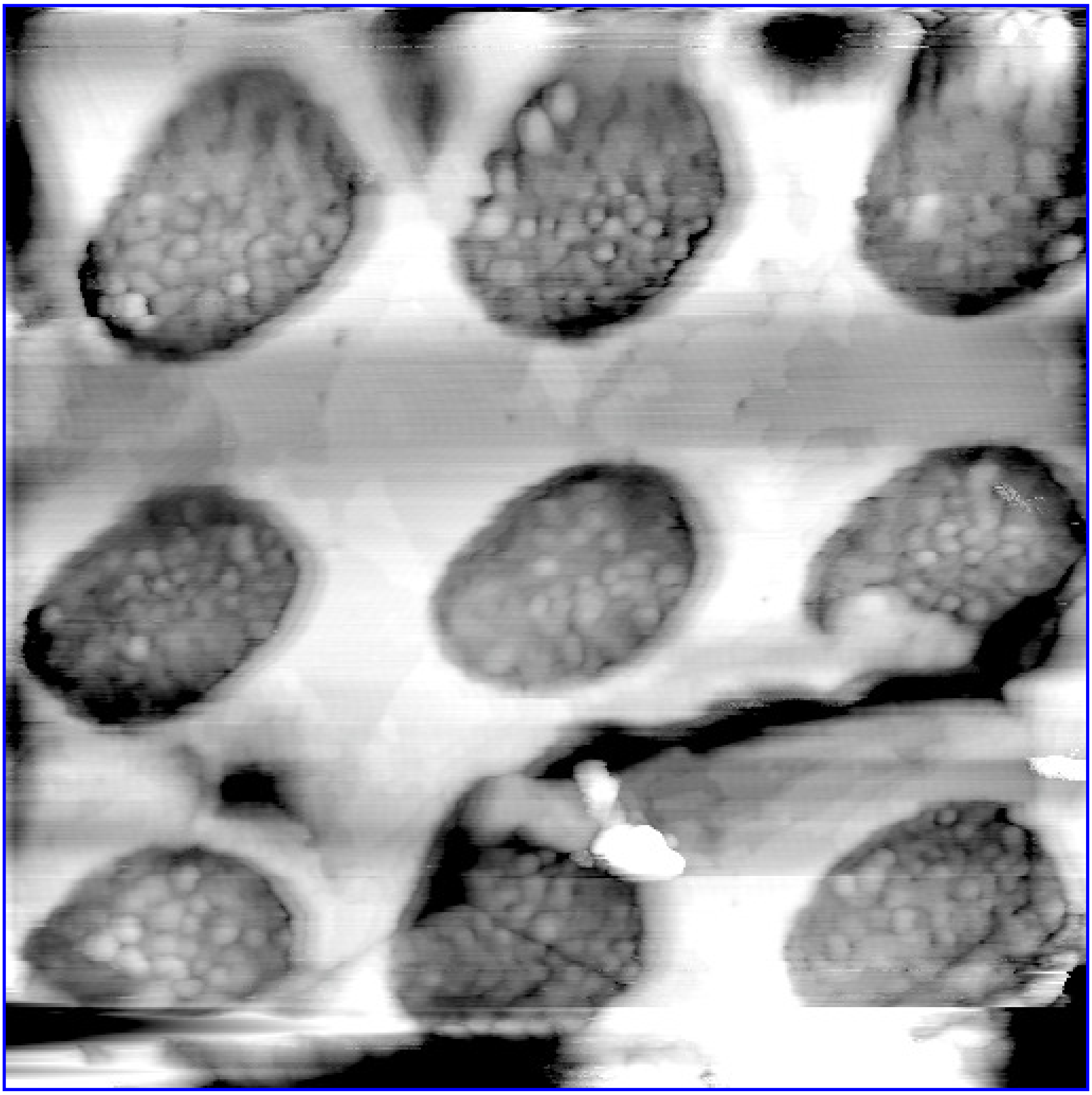

FIG. S4: **Nanopatterned $SiO_2$/Au(111):** The image has been measured in ACSTM mode and has a size of 867 x 867 nm$^2$.

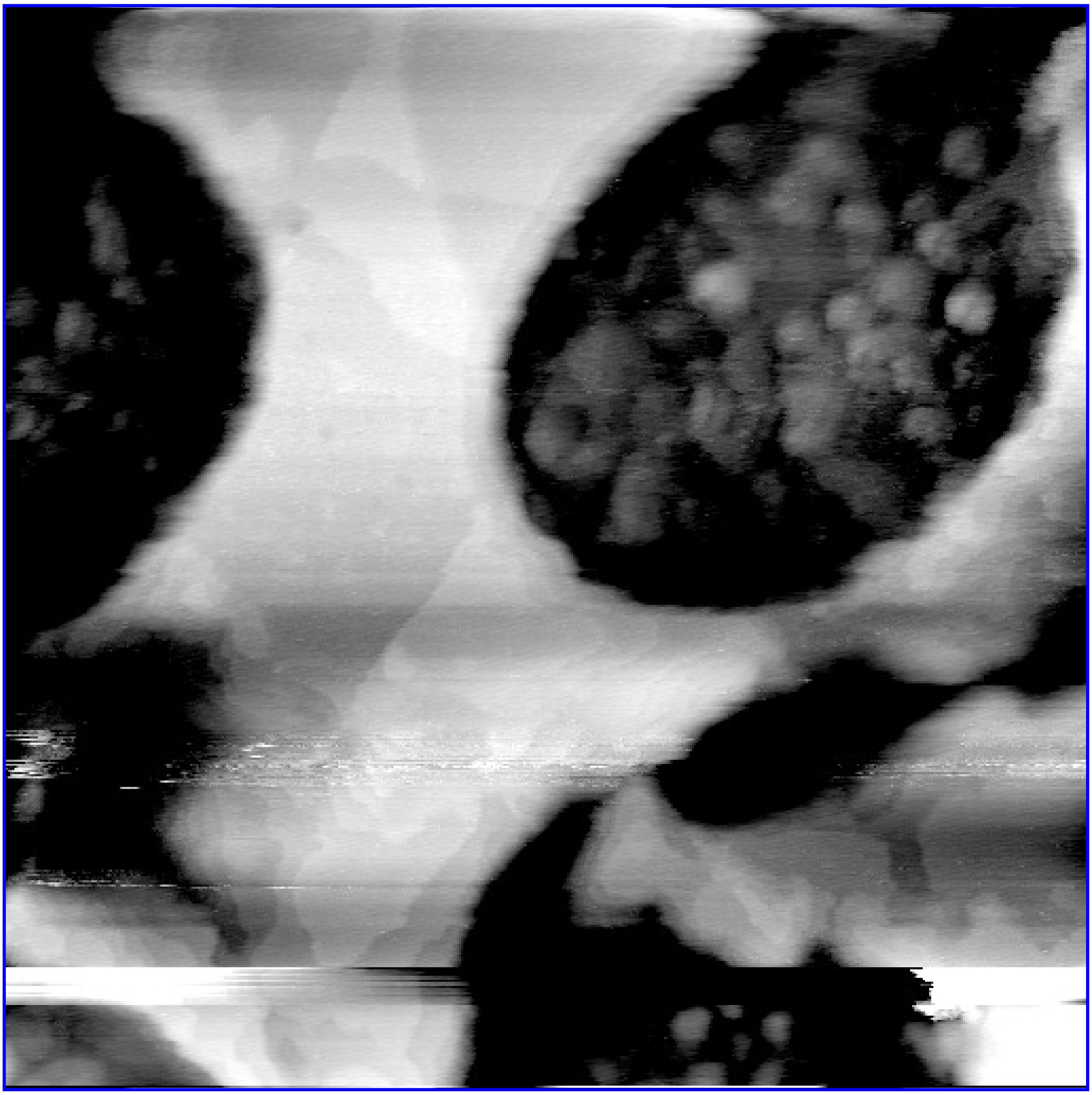

FIG. S5: **Nanopatterned $SiO_2$/Au(111):** The image has been measured in ACSTM mode and has a size of 434 x 434 $nm^2$. The particular brightness and contrast are chosen to enhance the visibility of the atomic steps on the oxide structure.

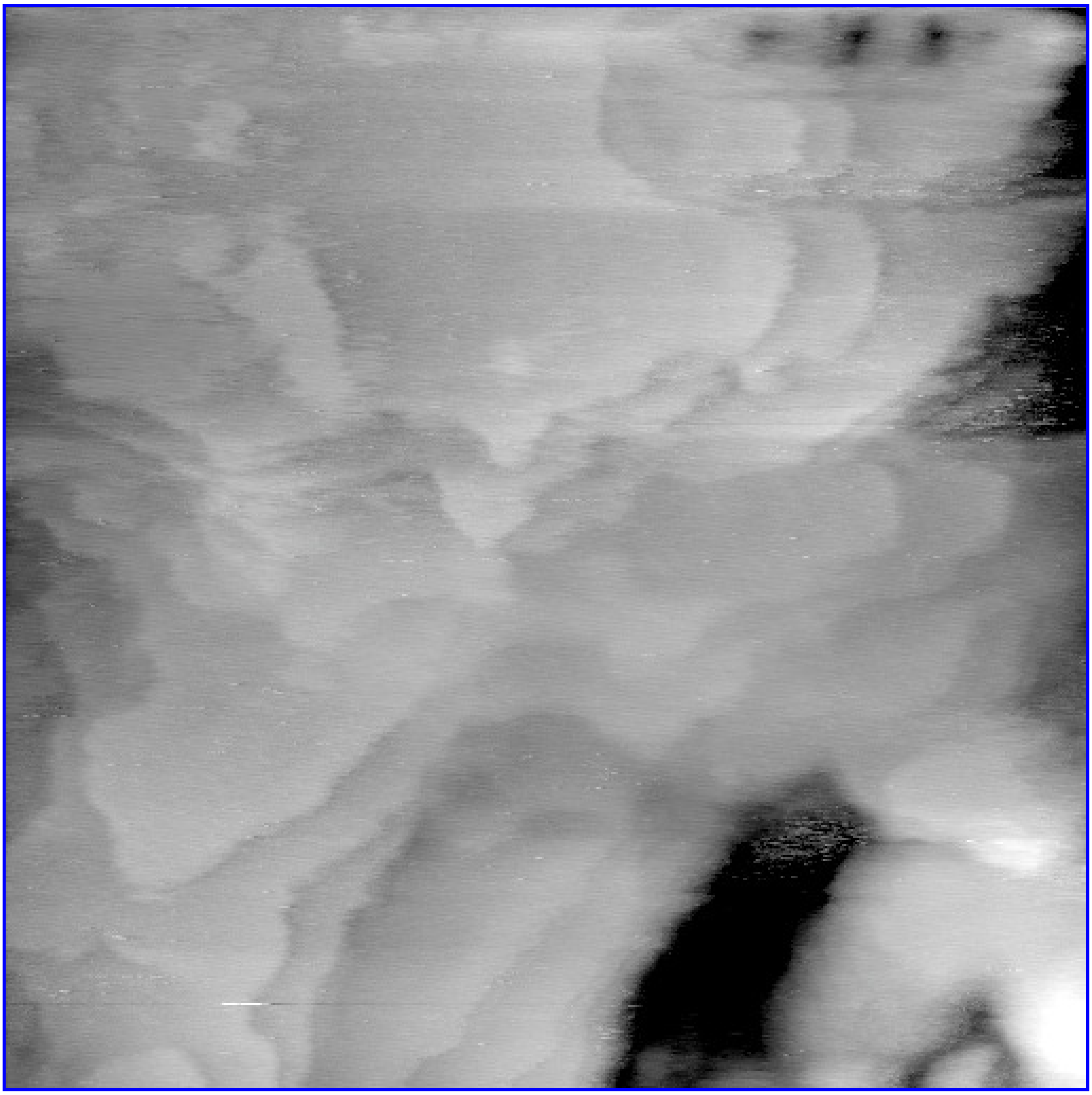

FIG. S6: **Nanopatterned $SiO_2$/Au(111):** The image has been measured in ACSTM mode and has a size of 217 x 217 nm$^2$.

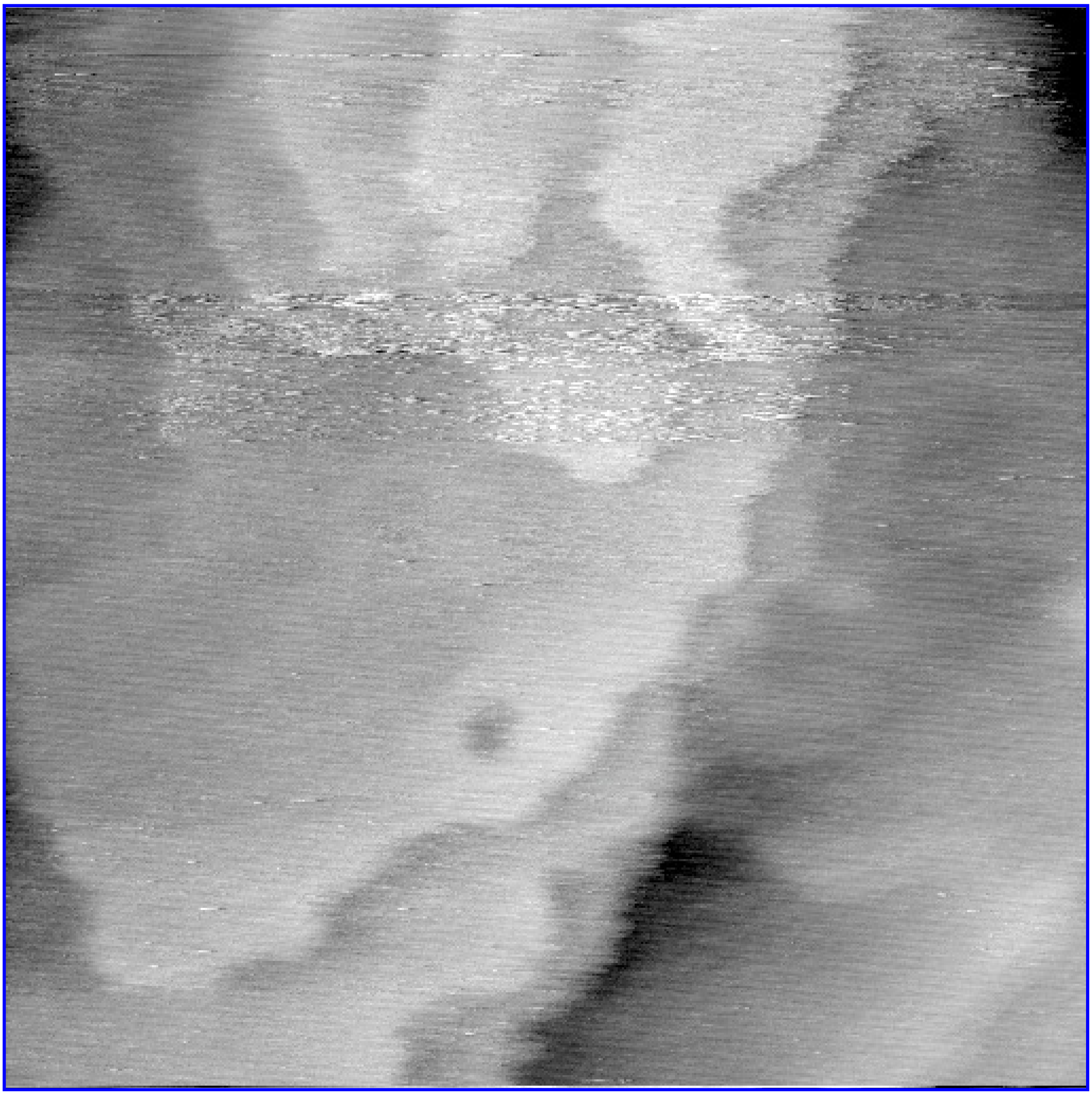

FIG. S7: **Nanopatterned $SiO_2$/Au(111):** The image has been measured in ACSTM mode and has a size of 106 x 106 $nm^2$.

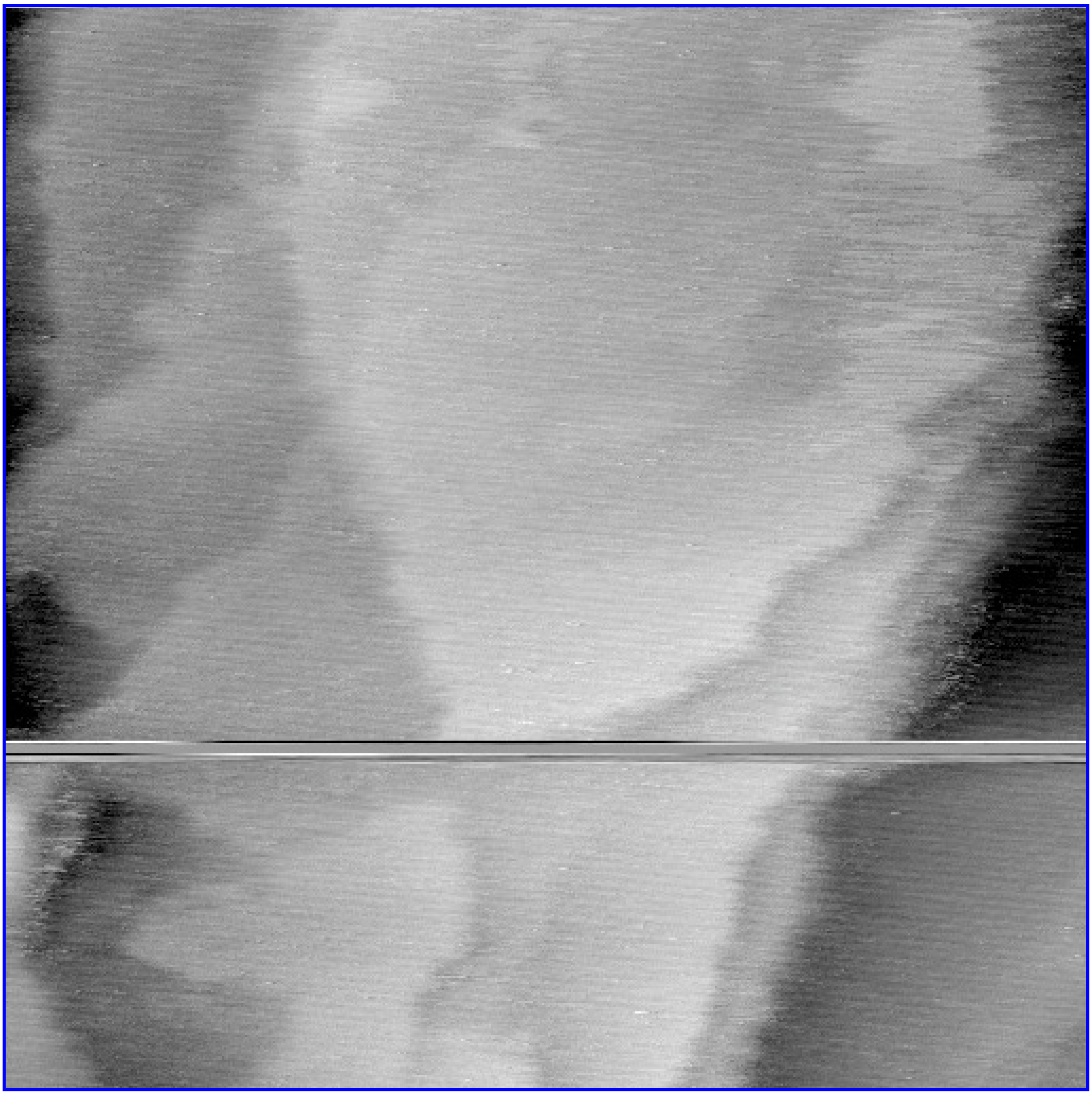

FIG. S8: **Nanopatterned $SiO_2$/Au(111):** The image has been measured in ACSTM mode and has a size of 106 x 106 nm$^2$.

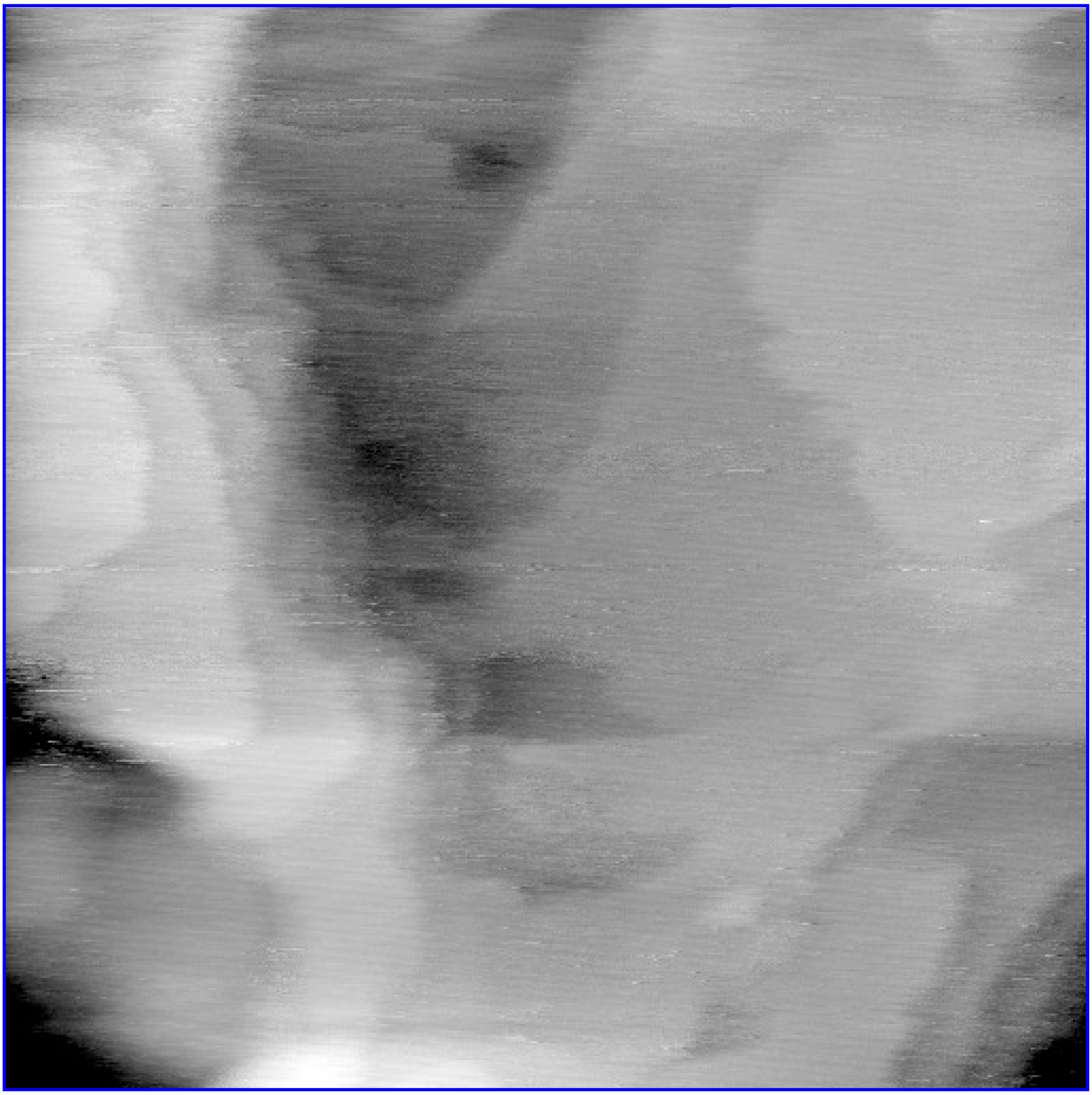

FIG. S9: **Nanopatterned $SiO_2$/Au(111):** The image has been measured in ACSTM mode and has a size of 106 x 106 $nm^2$.

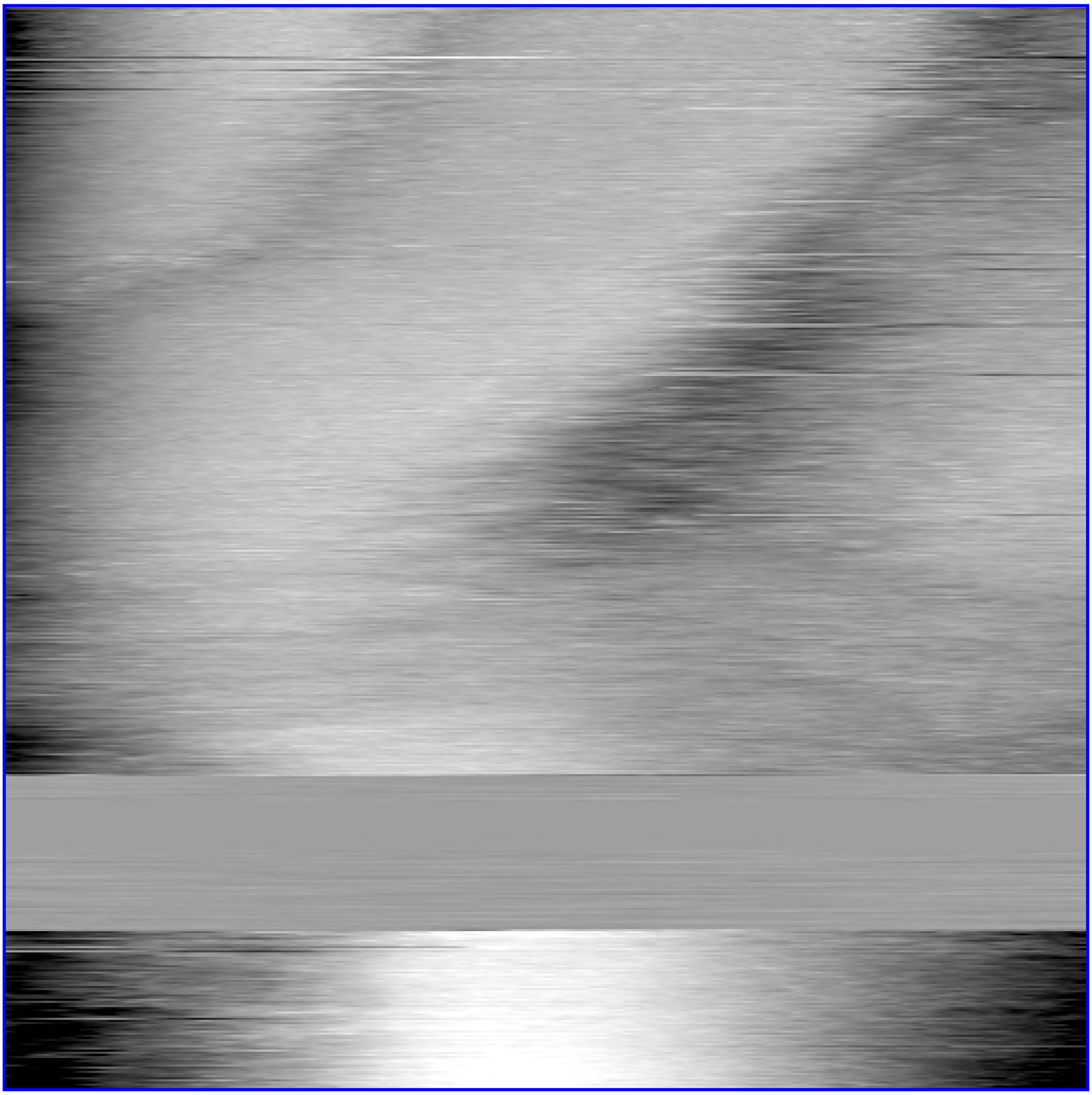

FIG. S10: **Nanopatterned $SiO_2$/Au(111):** The image has been measured in ACSTM mode and has a size of 31 x 31 $nm^2$.

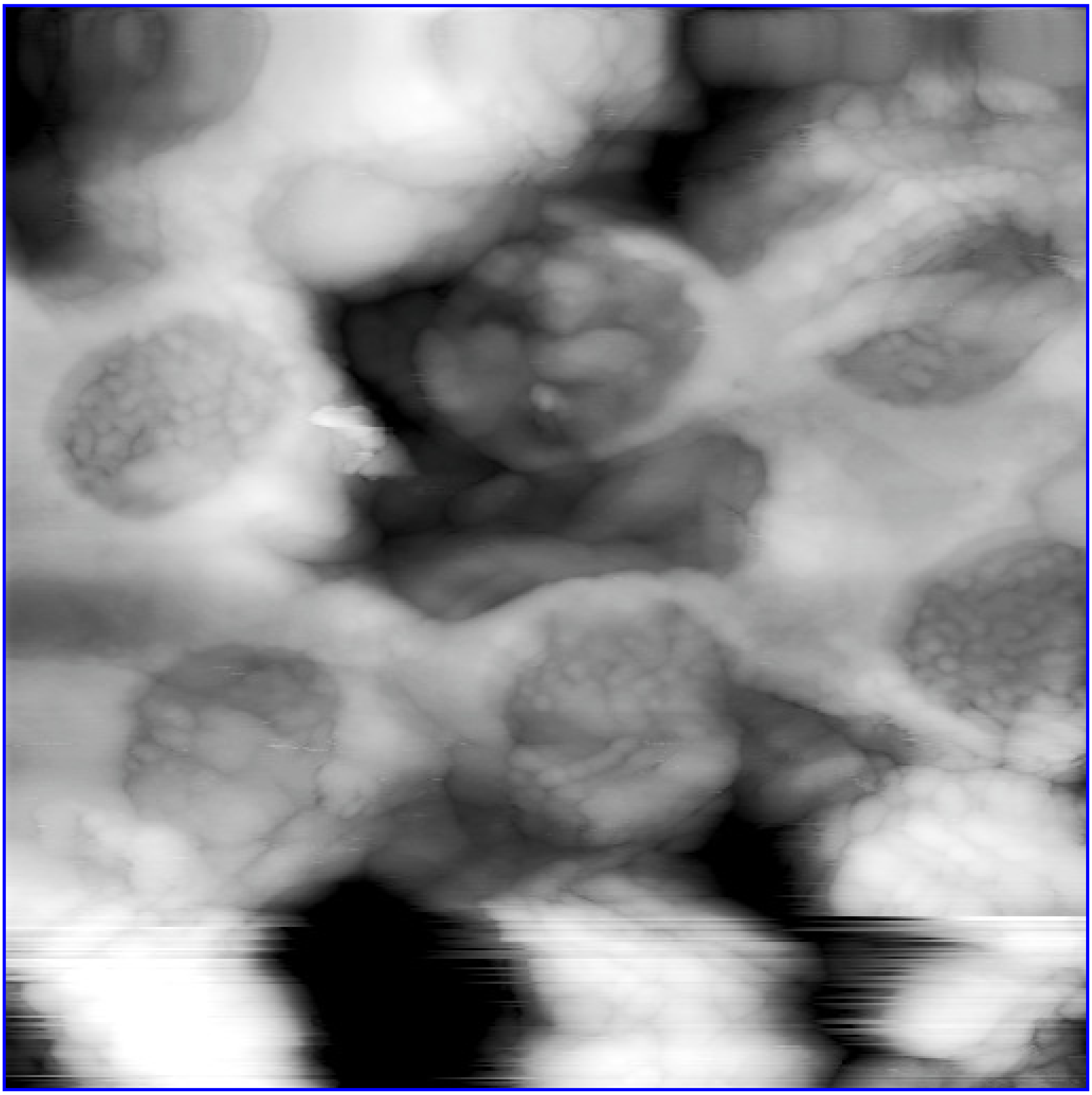

FIG. S11: **Nanopatterned $SiO_2$/Au(111):** The image has been measured in ACSTM mode and has a size of 867 x 867 $nm^2$. This image is measured on a different day with a different tip. Image artifacts of a multiple tip are visible such that a new tip was fabricated for further measurements.

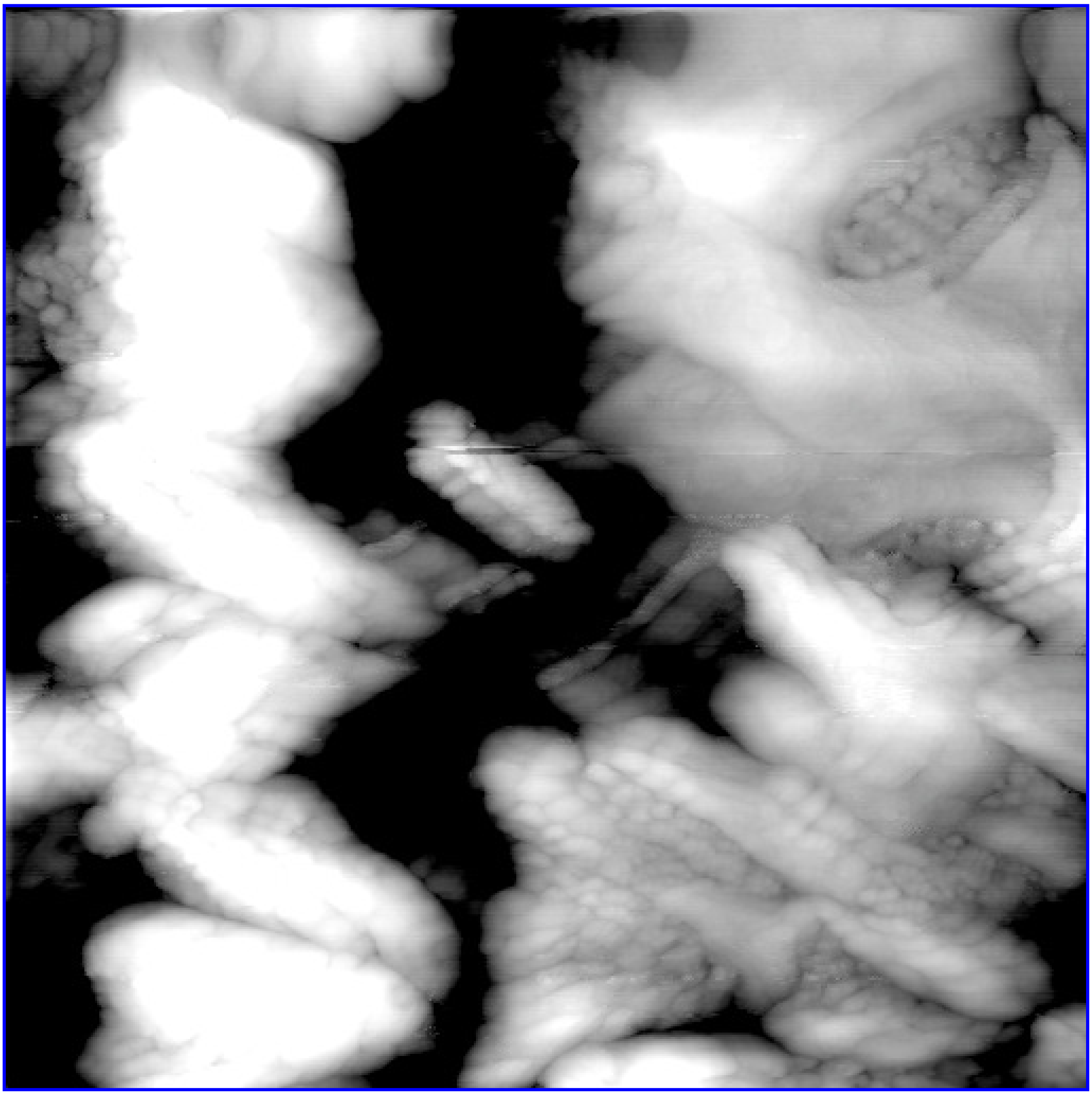

FIG. S12: **Nanopatterned $SiO_2$/Au(111):** The image has been measured in ACSTM mode and has a size of 867 x 867 $nm^2$. The particular brightness and contrast are chosen to enhance the visibility of the atomic steps on the oxide structure. This image is measured again on a different day with a different tip. And again, image artifacts of a multiple tip are visible such that a new tip was fabricated for further measurements.

[1] G.P. Kochanski, Phys. Rev. Lett. **62** 2285 (1989).

[2] S.J. Stranick and P. S. Weiss, J. Phys. Chem. **98** 1762 (1994).

[3] S.J. Stranick and P. S. Weiss, Rev. Sci. Instrum. **65** 918 (1994).

[4] L.A. Bumm and P.S. Weiss, Rev. Sci. Instrum. **66** 4140 (1995).

[5] U. Kemiktarak, T. Ndukum, K.C. Schwab, and K.L. Ekinci, Nature **450** 85 (2007).

[6] E. Matsuyama, T. Kondo, H. Oigawa, D. Guo, S. Nemoto, and J. Nakamura, Nature Sci. Rep. **4** 6711 (2014).

[7] M. Hervé, M. Peter, and W. Wulfhekel, Appl. Phys. Lett. **107** 093101 (2015).

[8] C. Saunus, J.R. Bindel, M. Pratzer, and M. Morgenstern, Appl. Phys. Lett. **102** 051601 (2013).

[9] I. Moult, M. Herve, and Y. Pennec, Appl. Phys. Lett. **98** 233103 (2011).

[10] M. Kasper, G. Gramse, J. Hoffmann, C. Gaquiere, R. Feger, A. Stelzer, J. Smoliner, and F. Kienberger, J. of Appl. Phys. **116** 184301 (2014).

[11] T.L. Cocker, D. Peller, P. Yu, J. Repp, and R. Huber, Nature **539** 263 (2016).

[12] M. Müller, N.M. Sabanés, T. Kampfrath, and M. Wolf, ACS Photonics **7** 2046 (2020).

[13] D.E. Steinhauer and S.M. Anlage, Appl. Phys. **89** 2314 (2001).

[14] T. Delattre, C. Feuillet-Palma, L.G. Herrmann, P. Morfin, J.-M. Berroir, G. Fève, B. Plaçais, D. C. Glattli, M.-S. Choi, C. Mora, and T. Kontos, Nature Phys. **5** 208 (2009).

[15] S.Loth, M. Etzkorn, C.P. Lutz, D. M. Eigler, and A.J. Heinrich, Science **329** 1628 (2010).

[16] E.J. Patiño, M. Aprili, M.G. Blamire, and Y. Maeno, Phys. Rev. B **87** 214514 (2013).

[17] K.M. Bastiaans, D. Chatzopoulos, J.-F. Ge, D. Cho, W.O. Tromp, J.M. van Ruitenbeek, M.H. Fischer, P.J. de Visser, D.J. Thoen, and M.P. Allan, Science **374** 608 (2021).

[18] M.E. Barber, E. Yue Ma, and Z.-X. Shen, Nature Rev. Phys. **4** 61 (2022).

[19] J. Schultheiß, T. Rojac, and D. Meier, Adv. Electron. Mater. **8** 2100996 (2021).

[20] J. Schmidt, D. H. Rapoport, and H.-J. Fröhlich, Rev. Sci. Instrum. **70** 3377 (1999).

[21] J.M. de Voogd, M.A. van Spronsen, F.E. Kalff, B. Bryant, O. Ostojić, A.M.J.den Haan, I.M.N. Groot, T.H. Oosterkamp, A.F. Otte, M.J. Rost, Ultramicroscopy **181** 61 (2017).

[22] M.J. Rost, M. Allan, and K. Heeck, Patent WO2025163175A1,

[23] S. Hellmüller, M. Pikulski, T. Müller, B. Küng, G. Puebla-Hellmann, A. Wallraff, M. Beck,

K. Ensslin, T. Ihn, Appl. Phys. Lett. **101** 042112 (2012).

[24] T.R. Albrecht, P. Grütter, D. Horne, and D. Rugar, J. of Appl. Phys. **69** 668 (1991).

[25] J.R. Matey and J. Blanc, J. of Appl. Phys. **57** 1437 (1985).

[26] L. Fumagalli, G. Ferrari, M. Sampietro, and G. Gomila, Appl. Phys. Lett. **91** 243110 (2007).

[27] C.Y. Nakakura, P. Tangyunyong, M.L. Anderson, (2007). *Scanning Capacitance Microscopy* in: S. Kalinin and A. Gruverman (eds) *Scanning Probe Microscopy* Springer, New York, NY (2007).

[28] G. Benstetter, R. Biberger, and D. Liu, Thin Solid Films **517** 5100 (2009).

[29] L. Fumagalli, G. Ferrari, M. Sampietro, I. Casuso, E. Martínez, J. Samitier, and G. Gomila, Nanotechnology **17** 4581 (2006).

[30] N.C. Bruce, A. García-Valenzuela, and D. Kouznetsov, J. Phys. D: Appl. Phys. **33** 2890 (2000).

[31] S. Lányi and M. Hruskovic, J. Phys. D: Appl. Phys. **36** 598 (2003)

[32] S.Y. Vasil'ev and A. V. Denisov, Technical Physics, **45,1** 99 (2000).

[33] T. Smith, J. Colloid Interface Sci. **75**, 51 (1980).

[34] D. Fragouli, T.N. Kitsopoulos, L. Chiodo, F. Della Sala, R. Cingolani, S.G. Ray, and R. Naaman, Langmuir **23**, 6156 (2007) and references therein

[35] M.J. Rost ”High-Speed Electrochemical STM” in “Encyclopedia of Interfacial Chemistry: Surface Science and Electrochemistry” K. Wandelt, Elsevier, pp. 180-198, (2018), ISBN: 9780128097397

[36] F. Simeone, D. Kolb, S. Venkatachalam, and T. Jacob, Angewandte Chemie International Edition **46**, 8903 (2007).

[37] W.-D. Schneider, Surf. Sci. **514** 74 (2002).

[38] D. Scholma and M.J. Rost, Thin Solid Films **835**, 140861 (2026).

[39] W. Steurer, A. Apfolter, M. Koch,T. Sarlat,E. Søndergård, W.E. Ernst, and B. Holst, Surf. Sci. **601** 4407 (2007).

[40] D. Scholma and M.J. Rost, in preparation

[41] https://www.vritratech.com/Quartz-Wafer.html

[42] Goodfellow, https://www.goodfellow.com/

[43] A. Halpern, *Schaum's Outlines Beginning Physics II*, New York: McGraw-Hill Companies, Inc., p. 141 (1998)

[44] R.W. Ward, *The Constants of Alpha Quartz*, IEEE Ultrasonics, Ferroelectrics and Frequency Control Society, September (1992).

[45] Y. Cui, S. Tosoni, W.-D. Schneider, G. Pacchioni, N. Nilius, and H.-J. Freund, Phys. Rev. Lett. **114** 016804 (2015).

[46] R.M. Feenstra, Phys. Rev. B **50**, 4561 (1994).

[47] G.J. de Raad, D.M. Bruls, P.M. Koenraad, and J.H. Wolter, Phys. Rev. B **66**, 195306 (2002).

[48] A. Revil and P.W.J. Glover, Geophys. Res. Lett. **25**, 691 (1998).

[49] H. Jain, Surf. Sci. **186**, 256 (1987).

[50] M. Tschapek, R. Santamaria, and I Natale, Electrochim. Acta **14**, 889 (1969).

[51] R. Guckenberger, M. Heim, G. Cevc, H.F. Knapp, W. Wiegräbe, and A. Hillebrand, Science **266** 1538 (1997).

[52] C.J.T. de Grotthuss, Ann. Chim. 58, 54 (1806).

[53] S. Cukierman, Biochimica et Biophysica Acta (BBA) - Bioenergetics, 1757, 876 (2006).

[54] J. Redondo et al. , Sci. Adv. **10** eadp7833 (2024).

[55] C. Roelcke, L.Z. Kastner, M. Graml, A. Biereder, J. Wilhelm, J. Repp, R. Huber, and Y.A. Gerasimenko, Nature Photonics **18**, 595 (2024).

[56] T. Choi, W. Paul, S. Rolf-Pissarczyk, A.J. Macdonald, F.D. Natterer, K. Yang, P. Willke, C.P. Lutz, and A.J. Heinrich, Nature Nanotechnology **12**,420 (2017).

[57] J. Siebrecht, H. Huang, P. Kot, R. Drost, C. Padurariu, B. Kubala, J. Ankerhold, J. Carlos Cuevas, and C.R. Ast, Nature Communications **14** 6794 (2023).

[58] S. Grall, I. Alić, E. Pavoni, M. Awadein, T. Fujii, S. Müllegger, M. Farina, N. Clément, G. Gramse, Small **17** 2101253 (2021).

[59] I. Battisti, G. Verdoes, K. van Oosten, K.M. Bastiaans, and M.P. Allan, Rev. Sci. Instrum. **89,12** 123705 (2018).

[60] When the tip is far from the surface, so no tunneling occurs, and the compensation capacitance is set to zero, the system impedance is determined solely by the stray capacitance and lies on the imaginary axis. Setting the compensation capacitance to the highest value leads to overcompensation and a sign reversal (180° degree phase shift). Optimal tuning occurs at 90° with a detectable minimum. Under these conditions, the current through the tip appears on the real axis.

[61] When tuning the compensation capacitance $C_{comp}$ on Au inside a hole without an active PLL,

the circuit becomes slightly detuned when measuring on the surrounding $SiO_2$.

## Acknowledgements

The author acknowledges K. Heeck for pointing out the advantages of a Balun and D. Scholma for his work in the NanoLab. Special acknowledgments go to Milan P. Allan, who supported this research not only with discussions, but also with financial means.

## Competing interests

The author declares no competing financial interests.